\documentclass[aps, twocolumn, amsmath]{revtex4}
\usepackage{dcolumn}% Align table columns on decimal point
\usepackage{bm}% bold math
\usepackage{graphicx}% Include figure files
\usepackage{lipsum}
\usepackage{color}
\usepackage{subfigure}
\usepackage{hyperref}
\usepackage{latexsym}
\usepackage{amsthm}
\usepackage{amssymb}
\usepackage{braket}
\DeclareGraphicsExtensions{.jpg,.pdf, .mps, .png, .eps, .ps, .EPS,.gif}

\begin{document}

\def\be{\begin{equation}}
\def\ee{\end{equation}}

\def\bc{\begin{center}}
\def\ec{\end{center}}
\def\bea{\begin{eqnarray}}
\def\eea{\end{eqnarray}}
\newcommand{\avg}[1]{\langle{#1}\rangle}
\newcommand{\Avg}[1]{\left\langle{#1}\right\rangle}

\def\ie{\textit{i.\,e.,}}
\def\etal{\textit{et al.}}
\def\m{\vec{m}}
\def\G{\mathcal{G}}

\newcommand{\red}[1]{{\bf\color{green}#1}}

% Figures 
% \includegraphics[width=2.5  in]{picL064pcfor53894043.eps} %{picpc} 
%   \includegraphics[width=2.5  in]{picL064pc2for53894043.eps}%//{picpc2} 
%  \includegraphics[width=2.5  in]{picL064pc2-7-1.eps} %{picpc2}   Fig 2a
%   \includegraphics[width=2.5  in]{picL064pc-7-1.eps}  %{picpc}   Fig 2b
%   \includegraphics[width=3.5  in]{mannaP2p} 
%  \includegraphics[width=3.5  in]{ScalingPlotManna}           Fig. 5
%   \includegraphics[width=3 in]{FigMannaPk.pdf} 
%   \includegraphics[width=3 in]{manna081219figurek1.eps}
 %   \includegraphics[width=3 in]{manna081219figurek2.eps}
 %   \includegraphics[width=3 in]{manna081219figurek3.eps}
 %   \includegraphics[width=3 in]{manna081219figurek4.eps}
                  
%Fig 3   Prpc.eps

%\newcommand{\rev}[1]{{\color{red}#1}}  %%% use this version to show revisions in red
\newcommand{\rev}[1]{#1}    %hide revisions

\title{Bond percolation between $k$ separated points on a square lattice}
%Percolation between $k$ separated points in two dimensions, \rev{for bond percolation on a square lattice}}

\author{S. S. Manna}
\affiliation{Satyendra Nath Bose National Centre for Basic Sciences, Block-JD, Sector-III, Salt Lake, Kolkata-700106, India}
\author{Robert M. Ziff\footnote{rziff@umich.edu} }
\affiliation{Center for the Study of Complex Systems and Department of Chemical Engineering, University of Michigan, Ann Arbor, Michigan 48109-2136, USA}

\begin{abstract}We consider a percolation process in which $k$ points separated by a distance proportional to system size $L$ simultaneously 
   connect together ($k>1$), or a single point at the center of a system connects to the 
   boundary ($k=1$), through adjacent connected points of a single cluster. These processes 
   yield new thresholds $\overline p_{ck}$ defined as the average value of $p$ at which the desired connections first occur. These thresholds are 
not sharp as the distribution of values of $p_{ck}$  for individual samples remains broad in the limit of 
   $L \to \infty$. We study $\overline p_{ck}$ for bond percolation on the square lattice, 
   and find that $\overline p_{ck}$ are above the normal percolation threshold $p_c = 1/2$ 
   and represent specific supercritical states. The $\overline p_{ck}$ can be related to 
   integrals over powers of the function $P_\infty(p)$ equal to the probability a point 
   is connected to the infinite cluster; we find numerically from both direct simulations 
   and from measurements of $P_\infty(p)$ on $L\times L$ systems that, for $L \to \infty$, 
   $\overline p_{c1} = 0.51755(5)$, $\overline p_{c2} = 0.53219(5)$, $\overline p_{c3} = 
   0.54456(5)$, and $\overline p_{c4} = 0.55527(5).$  The percolation thresholds $\overline p_{ck}$
   remain the same, even when the $k$ points are randomly selected within the lattice. We show 
   that the finite-size corrections scale as $L^{-1/\nu_k}$ where $\nu_k = \nu/(k \beta +1)$, 
   with $\beta=5/36$ and $\nu=4/3$ being the ordinary percolation critical exponents, so 
   that $\nu_1= 48/41$, $\nu_2 = 24/23$, $\nu_3 = 16/17$, $\nu_4 = 6/7$, etc. We also study 
   three-point correlations in the system, and show how for $p>p_c$, the correlation ratio goes to
   1 (no net correlation)  as $L \to \infty$, while at $p_c$ it reaches the known value of 1.022.
\end{abstract}

\maketitle

\section{Introduction}
\label{sec:introduction}

Percolation is the study of long-range connectiveness in systems such as  graphs or lattices in which the sites 
or bonds are randomly occupied with probability $p$. There is a well-defined threshold $p_c$ at which the average 
size of a cluster first becomes infinite. The threshold can also be defined by considering finite systems (say an 
$L \times L$ square), and studying the probability that a single cluster connects or spans two opposite sides. The 
average value of $p$ at which spanning first occurs yields an estimate for $p_c(L)$, and using finite-size scaling 
one can predict the value of $p_c$ for $L \to \infty$.   In this  case, the threshold is sharp as $L \to \infty$.  For a square lattice with bond percolation, for example, 
one has $p_c = 1/2$ \cite{StaufferAharony94,ReynoldsStanleyKlein80}.

\rev{Percolation has received a great deal of 
attention over the years; some recent papers include a study of regular and inverse percolation of rigid rods \cite{RamirezCentresRamirezPastor18},
continuum percolation of overlapping polyhedra \cite{XuZhuJiangJiao19},
percolation over varied ranges of transmission \cite{KunduManna},
percolation on a distorted lattice 
\cite{MitraSahaSensharma19},
percolation of $k$-mers undergoing random sequential adsorption \cite{SlutskiiBarashYuTarasevich18}, 
percolation disassortativity on random networks \cite{MizutakaHasegawa18},
percolation for random sequential adsorption with relaxation \cite{KunduAraujoManna18}, percolation over a range of interactions \cite{OuyangDengBlote18},
percolation in high dimensions and on a random graph \cite{HuangHouWangZiffDeng18}, 
percolation on hypercubic lattices in high dimensions \cite{MertensMoore18,MertensMoore18b},
percolation of  the elastic backbone \cite{FilhoAndradeHerrmannMoreira18},
universality in explosive percolation \cite{SabbirHassan18}, 
crossing probabilities for polygons \cite{FloresSimmonsKlebanZiff17},
rigorous bounds for percolation thresholds \cite{Wierman17},
percolation on random jammed sphere packings \cite{ZiffTorquato17}, and percolation on hyperbolic manifolds \cite{KryvenZiffBianconi19}.  Clearly, percolation remains a very active field.}

For the ordinary percolation problem in $d$ dimensions, the connectivity is usually considered between the pair of opposite 
$(d-1)$ dimensional hypersurfaces. Naturally, the question arises, what happens if the connectivity is considered 
between the $(d-2)$, $(d-3)$, .... dimensional hypersurfaces? In this paper, we try with the simplest possible situation, 
that is the connectivity between the $(d-2)$ dimensional hypersurfaces in $d = 2$. More specifically, we study the 
percolation problem between the $k$ widely separated points (dimension 0) on the two-dimensional square lattice, or between
a single point and the boundary of the system.

The first threshold we consider is defined as the average value of $p$ at which a point in the center of a square 
system first connects to any point on the boundary.  This defines the threshold $\overline p_{c1}$.  The other 
thresholds are defined as the average value of $p$ at which $k$ points separated far apart in a periodic system all 
first connect; we call those thresholds $\overline p_{ck}$.   These thresholds are all greater than $p_c$, indicating 
that we are in the supercritical regime of percolation where there is a percolating net throughout the system. Being 
in a supercritical state is expected since connecting a large  cluster to a specific single point at the normal 
critical point $p_c$  occurs with low probability (unlike connecting to a boundary, for example, which can occur 
through many paths and is much easier).  Connecting to a boundary is a universal property that survives at the 
critical point when the lattice spacing goes to zero, while in that limit the probability of connecting to a 
single point goes to zero.  When going to the supercritical regime, the probability of connecting to a point 
can be raised to a significant value, and this allows different points to connect together simultaneously with 
a sufficient probability to be observed.   

We carried out  computer simulations to find the values of $\overline p_{ck}$ directly for $k = 1, 2, 3$ and 4.  
We also developed a theory to connect $\overline p_{ck}$ to $P_\infty$, the percolation function that gives the 
probability a given point belongs to the infinite cluster, or the largest cluster for a finite system.  By directly 
simulating $P_\infty$ for this system, we are able to verify numerically that the relation to $\overline p_{ck}$ 
is valid.  The analysis also shows that, unlike in the case of the usual percolation threshold, the distribution 
of $\overline p_{ck}$ for individual systems is broad and does not become sharp as the system size goes to infinity. 
That is, there are large fluctuations in the states of these systems defined by these percolation criteria.

\begin{figure}[htbp] %  figure placement: here, top, bottom, or page
   \centering
   \includegraphics[width=3.0 in]{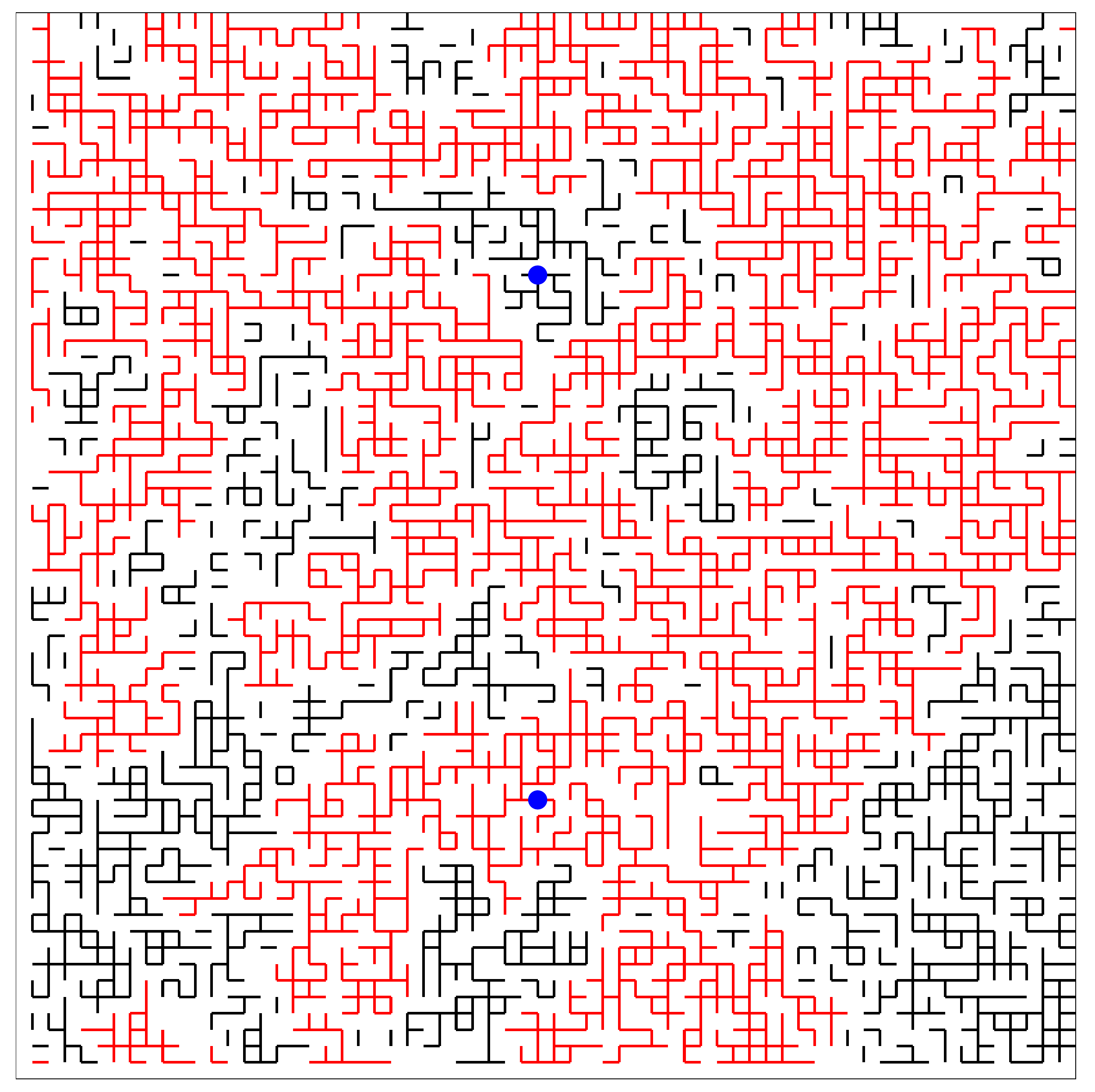} \\ %{picpc}
   (a) \\
   \includegraphics[width=3.0 in]{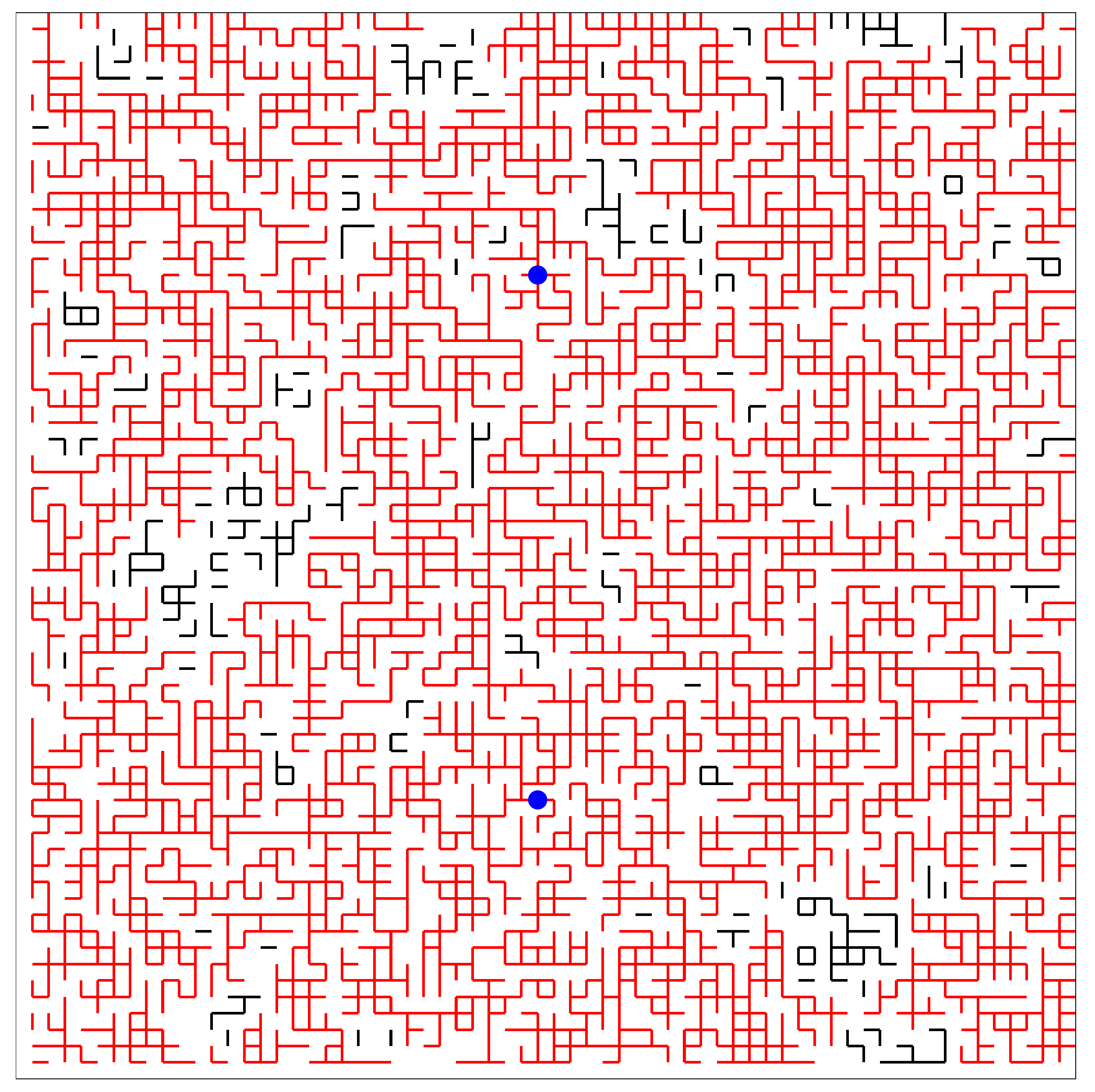} \\ %{picpc2} 
   (b) \\
   \caption{
   Two illustrations of the system are presented for a lattice of size $64\times64$ 
   with periodic boundary conditions, and with $k = 2$ anchor points (marked by filled 
   blue circles) separated by a distance of $32$ lattice units. Bonds of the largest 
   occupied cluster are shown in red \rev{(grey)}, and all other occupied bonds are shown in black.  
   (a) A system where the number of bonds is exactly 4096 or $p = 4096/8192= 1/2 = p_c$, 
   without connection between the two anchor points. (b) The same system where the 
   number of occupied bonds is increased to 4415 bonds or $p_{c2} = 4415/8192 \approx 0.53894$, at 
   which point connection between the anchors first occurred. This is a typical example 
   where the threshold is near the average value of $\overline p_{c2} = 0.5322$ and shows that in 
   this supercritical regime there is a percolating network that goes essentially 
   throughout the entire system.
   } \label{fig:pic}
\end{figure}

In Fig.\ \ref{fig:pic} we show pictures of simulations of a $64 \times 64$ periodic system in which the first 
connection between the two anchor points occurred when 4415 bonds were placed down, or at $p_{c2} = 4415/8192 \approx 0.53894$, 
and the same system at the standard threshold $p = 4096/8192 = 1/2 = p_c$, at which point no connection exists 
between the two anchor points for this system. The value of $p_{c2}$ for this sample is close to the average 
value $\overline p_{c2} = 0.5312$ found by averaging over many realizations. It can be seen that, at $p_{c2}$, there is one overwhelming ``infinite''
cluster throughout the system,  and finite clusters are very small.  This behavior illustrates the idea behind 
our conjecture that in the supercritical region, the probability that $k$ points are connected together is equal 
to $[P_\infty(p)]^k$.

\begin{figure}[htbp] %  figure placement: here, top, bottom, or page
   \centering
   \includegraphics[width=3.0 in]{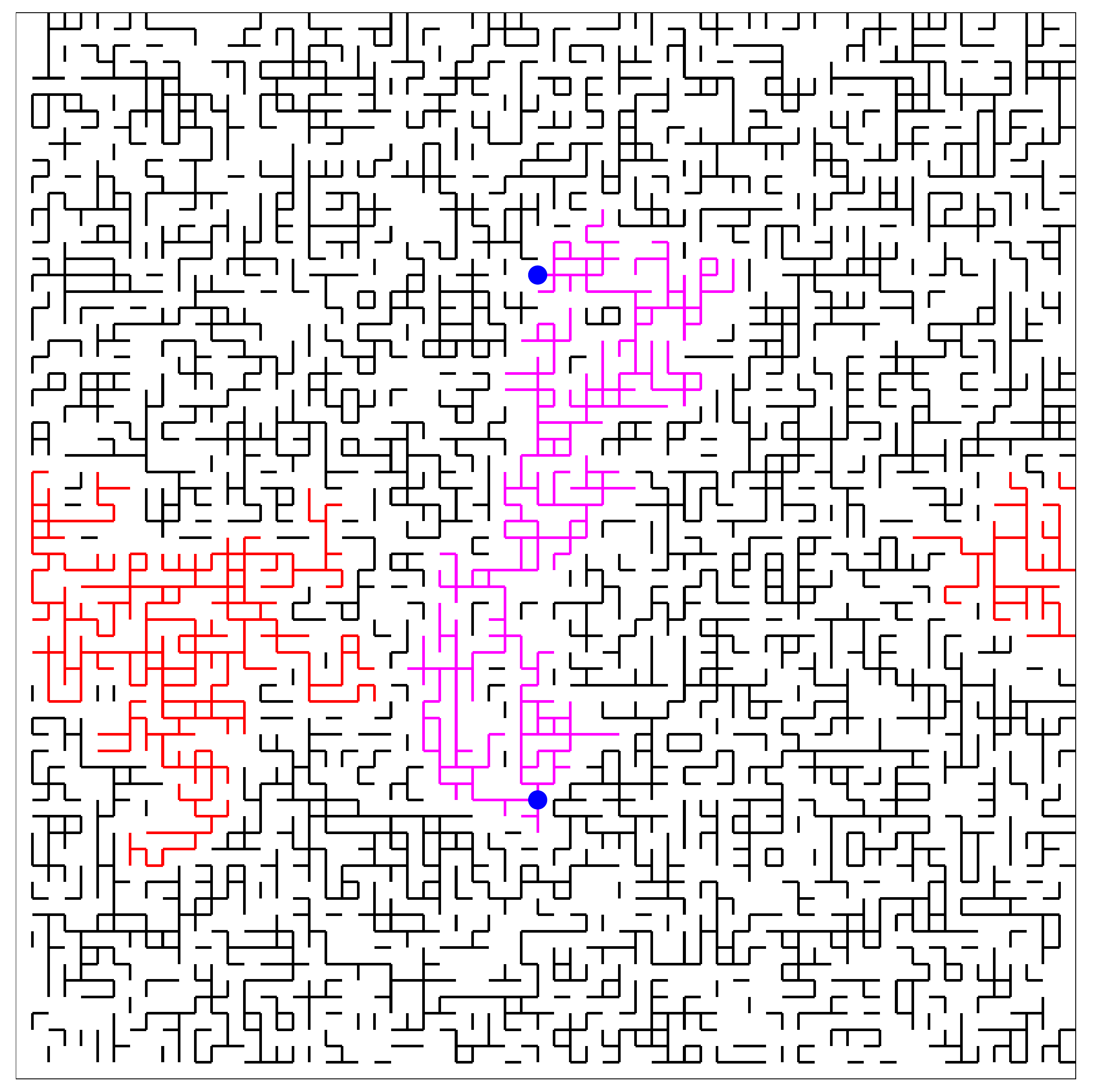} \\ %{picpc2} 
   (a) \\
   \includegraphics[width=3.0 in]{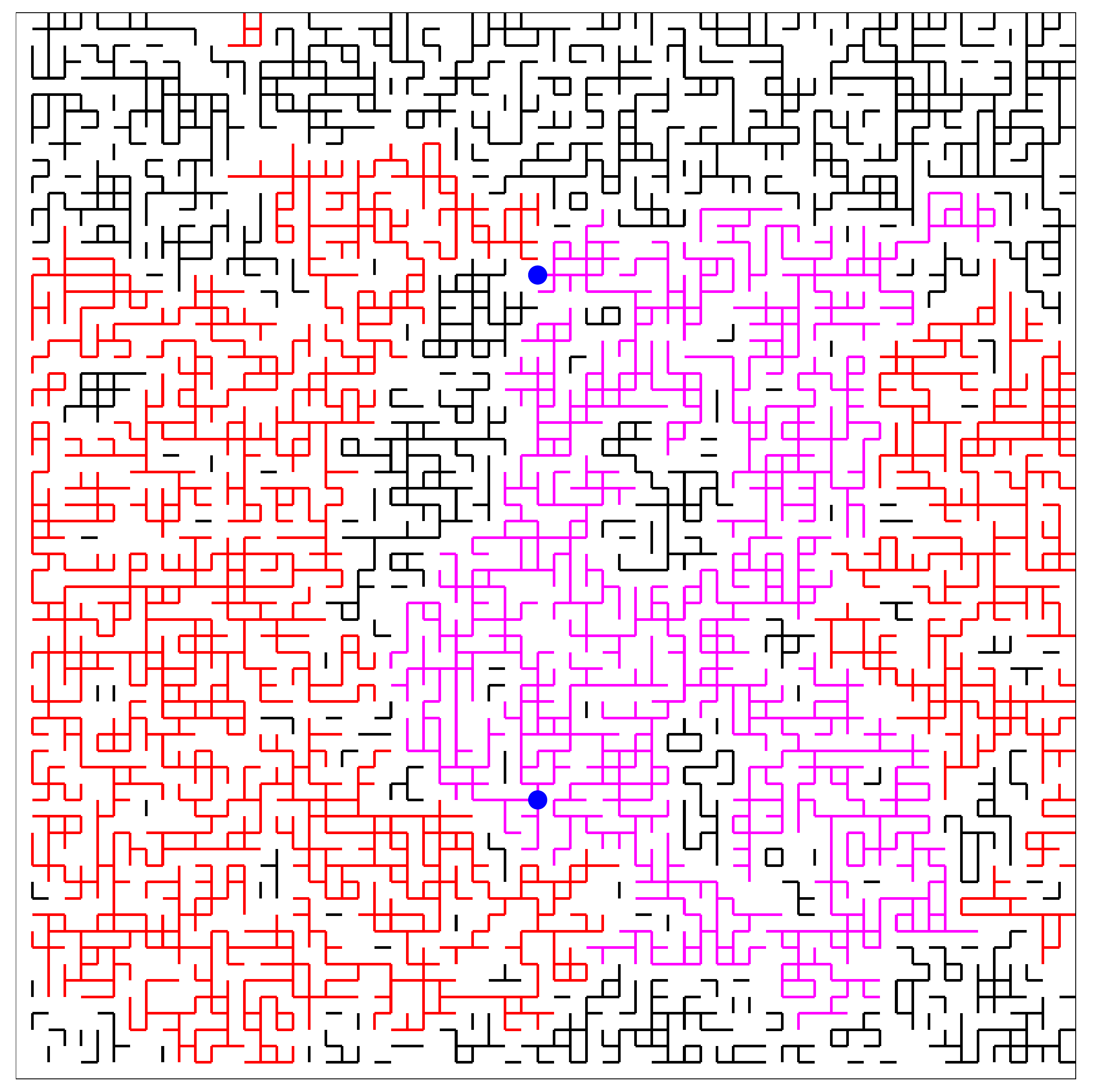} \\ %{picpc} 
   (b) \\   
   \caption{
   (a) Here the density of occupied bonds at the first connection occurs at $p = p_{c2} = 
   3660/8192= 0.44678$. (b) The same system where $p$ is increased to $p_c = 1/2$ is shown. 
   This is a very rare system in which the connection between the anchor points first 
   occurs substantially below $p_c$, where the point connecting cluster \rev{(magenta or light grey bonds)}
   is different from the largest cluster \rev{(red or grey bonds)}. In fact, the spanning cluster is 
   relatively small and does not extend over the whole system. Such behavior where spanning 
   occurs below $p_c$ can only happen in smaller systems. In most cases, the individual 
   values of $p_{c2}$ are larger than $p_c$ and the cluster connecting them is the 
   ``infinite" cluster that spreads over virtually the entire system \rev{as in Fig.\ \ref{fig:pic}(b).}
   }
   \label{fig:pic2}
\end{figure}

In Fig.\ \ref{fig:pic2} we show a very rare case where the connection between the anchor points occurred at a 
value substantially below $p = 1/2$; for large systems such cases appear with very low probability.

      We also studied a ratio involving three-point correlations and two-point correlations, 
   and show how that varies with the separation of the points compared with the size of the 
   system. This ratio has been studied previously at the critical point only 
   \cite{SimmonsZiffKleban09,DelfinoViti10}; here we study it for all $p$.

      In section \ref{sec:theory} we develop our theory for $\overline p_{ck}$, including 
    the scaling of the estimates. In section 
   \ref{sec:methods} we describe our simulation methods, and in section \ref{sec:results} 
   we give the results of our simulations. In section \ref{sec:correlations} we consider 
   the problem of the three-point correlation ratio.  In section \ref{sec:discussion} we 
   discuss our results and give our conclusions.

\section{Theoretical analysis} 
\label{sec:theory}

      Here we develop a theory to predict $\overline p_{ck}$ from $P_\infty(p)$, and develop a scaling analysis 
   that allows one to predict the convergence exponents for the $\overline p_{ck}$.

\subsection{Relation to $P_\infty$}

     The first assumption is that we must be in the supercritical state, since only then will the $k$ points be 
  able to connect together via the infinite network.  At $p_c$, the infinite cluster is tenuous and fractal, 
  and does not connect to given points with a significant probability (for a large system), and below $p_c$ the clusters are all 
  small and it would be virtually impossible for points far apart to connect together.

Thus, for $k$ widely separated points to be all connected together, we hypothesize that they \rev{must be}  part of the infinite cluster 
in the supercritical state.   The probability a single point belongs to the infinite cluster is denoted as $P_\infty(p)$; 
for a finite system we can define $P_\infty(p,L) = s_\mathrm{max}/L^2$ where $s_\mathrm{max}$ is the number of sites in 
the largest cluster in the system.  Thus, we conjecture that the probability that $k$ widely separated points are connected must be equal to 
$[P_\infty(p)]^k$.  The probability \rev{density} that they first connect when the occupation probability is $p$ is then 
\begin{equation} \rev{P_r =}(d/dp)[P_\infty(p)]^k = k[P_\infty(p)]^{k-1} P_\infty'(p)
\label{eq:Pransatz}
\end{equation}
and the average value of $p$ at which the $k$ 
points first connect will be given by
\begin{equation}
\overline p_{ck} = \langle p \rangle =  \int_0^1 p (d/dp) [P_\infty(p)]^k  dp
\label{eq:pck1}
\end{equation}
Integrating by parts, we find
\begin{equation}
\overline p_{ck} = 1 - \int_0^1 [P_\infty(p)]^k  dp = \int_0^1 (1-[P_\infty(p)]^k)  dp
\label{eq:pck}
\end{equation}

For the problem of a single site connected to the boundary (corresponding to $\overline p_{c1}$), the above 
formulas also apply, taking $k = 1$.  In this case, the largest cluster surely connects to the boundary, so 
we are asking for just the probability that a point connects to the largest cluster, which is given by $P_\infty(p)$.   \rev{Note, for the case of $k = 1$, we do not use periodic boundary conditions.}

For $k>1$ the value of $\overline p_{ck}$ should be independent of the exact configuration of the $k$ points, 
as long as their relative distances grow with $L$, so that they become infinitely far apart as $L \to \infty$ 
and greater than the correlation length $\xi$, which is finite for any given $p > p_c$.
For finite systems, the specific configuration of the points will be relevant for the precise threshold.

We can make a very useful approximation for calculating $\overline p_{ck}$ from $P_\infty(p)$ for finite systems 
by simply assuming $P_\infty(p)=0$ for $p < p_c$, which is true for an infinite system. Then the integrand in 
the second form of equation (\ref{eq:pck}) is exactly 1 in the interval $0 < p < p_c$, and we can write as an 
alternative to (\ref{eq:pck})
\begin{equation}
\overline p_{ck} = p_c + \int_{p_c}^1 (1-[P_\infty(p)]^k)  dp
\label{eq:pckalt}
\end{equation}
\rev{where $p_c = 1/2$ for bond percolation on the square lattice.}
Equations (\ref{eq:pck}) and (\ref{eq:pckalt}) are identical when $L \to \infty$, but it will turn out that 
(\ref{eq:pckalt}) gives a much better estimate of $\overline p_{ck}$ for finite $L$.

\subsection{Scaling of the estimates}

If we assume that the mapping of our problem to $[P_\infty(p)]^k$ is correct for finite systems 
characterized by $P_\infty(p,L)$,  we can then estimate the scaling behavior of the estimates 
from finite-size scaling theory.  That theory states that for  $L \to \infty$ and  $p - p_c \to 0$ 
with $(p-p_c) L^{1/\nu}$ constant,
\begin{equation}
P_\infty(p,L) \sim a L^{-\beta/\nu} F(b(p-p_c) L^{1/\nu})
\label{scaling}
\end{equation}
where $a$ and $b$ are system-dependent constants (``metric factors") while $\beta$, $\nu$ and $F(z)$ 
are universal quantities, having the same values and behavior for all systems of a given dimensionality, 
and also a given system shape for the case of $F(z)$.  For $d = 2$, one has $\beta = 5/36$ and $\nu = 4/3$ 
\cite{StaufferAharony94}. 

We will apply this \rev{scaling} to the estimate for $\overline p_{ck}$ given by Eq.\ (\ref{eq:pck}).
First we  consider the  interval $p = (0,p_c)$.  In this interval, we assume that the finite-size 
effects are essentially those given by the scaling function $F(z)$, because when $p<p_c$, 
$P_\infty(p,\infty) = 0$.  That is, we assume the non-scaling corrections are unimportant for 
large $L$ for $p<p_c$.

Putting (\ref{scaling}) into the integral in Eq.\ (\ref{eq:pck}) over the interval $p = (0,p_c)$, we find
\begin{equation}
\int_0^{p_c} [P_\infty(p)]^k  dp  = a^k \int_0^{p_c} L^{-k \beta/\nu} [F(b(p-p_c)L^{1/\nu})]^k dp 
\end{equation}
and a change of variables yields 
\begin{eqnarray}
 \int_0^{p_c} [P_\infty(p)]^k  dp  &=&  a^k b^{-1} L^{-(k \beta+1)/\nu}  \int_{-b p_c L^{1/\nu}}^{0}  [F(z)]^k dz \cr \cr
 &\approx&  a^k b^{-1}  L^{-(k \beta+1)/\nu}  \int_{-\infty}^{0}  [F(z)]^k dz
 \label{eq:scaling2}
\end{eqnarray}
where $z = b(p-p_c) L^{1/\nu}$.  In the second integral in (\ref{eq:scaling2}) we extended the lower limit to 
$-\infty$, valid for large $L$ because the integrand decays exponentially for negative $z$.

Therefore, this contribution to the integral in (\ref{eq:pck}) should scale as $L^{-1/\nu_k}$ with
\begin{equation} 1/\nu_k = (k \beta + 1)/\nu = \frac{36+5k}{48} \label{eq:nuk}\end{equation}
 so that $1/\nu_1 = 41/48 = 0.854166$ and $1/\nu_2 =  23/24 = 0.958333$ etc.  

For $p > p_c$, it is not clear how to attack the finite-size corrections of the integral in (\ref{eq:pck}) because 
there are large non-scaling contributions to $P_\infty$ whose behavior we do not know, but it seems reasonable to 
assume that the finite-size corrections for $p>p_c$ scale the same as those we found for $p < p_c$, so we conjecture 
that the exponents $\nu_k$ above should characterize the full finite-size corrections to $\overline p_{ck}$.  \rev{That is,  we conjecture
\begin{equation}
\overline p_{ck}(L) = \overline p_{ck} + c L^{-1/\nu_k}
\label{eq:pckscaling}
\end{equation}
where $c$ is a constant and $\nu_k$ is given by Eq.\ (\ref{eq:nuk}).  The constant term
on the right-hand side, $\overline p_{ck}$, derives from the non-scaling parts of $P_\infty$ for $p > p_c$.

%The first term in the right hand side above ($\overline p_{ck}$) must follow from the non-scaling part of the integrand in Eq.\ \ref{eq:pckalt}, because the scaling part gives the finite-size correction.}

%  [This related to k = 2 only (???)   updated it to k]  
 %     Thus, we conclude that the \rev{integral of the } scaling part of $k [P_\infty(p)]^{k-1} P'_\infty(p)$ goes to zero as $L^{-(k \beta + 1)/\nu}$ as $L \to \infty$.  \rev{The non-scaling part presumably yields $p_{ck}-p_c$, when put in the integral in Eq.\ (\ref{eq:pck}), although we cannot show that explicitly.  In any case, we come to the conclusion that $p_{ck}$ scales as Eq.\ \ref{eq:  }.}
 
\rev{Note that it also follows from the scaling arguments above that $P_r = k [P_\infty(p)]^{k-1} P'_\infty(p)$ behaves with $L$ in the scaling regime as 
 \begin{eqnarray}
 P_r &\sim&  k a^{k-1}  L^{-(k-1) \beta/\nu} [F(b(p-p_c)L^{1/\nu})]^{k-1}\cr
 &\times& a L^{-\beta/\nu}F'(b(p-p_c)L^{1/\nu})  b L^{1/\nu} \cr 
 &\sim&  L^{ -(k \beta - 1)/\nu} G(b(p-p_c)L^{1/\nu})
 \label{eq:Prscaling}
  \end{eqnarray}
  }
}

%We will verify this conjecture numerically.

%-----------------------------------------------------------------------------------
\begin{figure}[t] %  figure placement: here, top, bottom, or page
   \centering
   \includegraphics[width=3.0in]{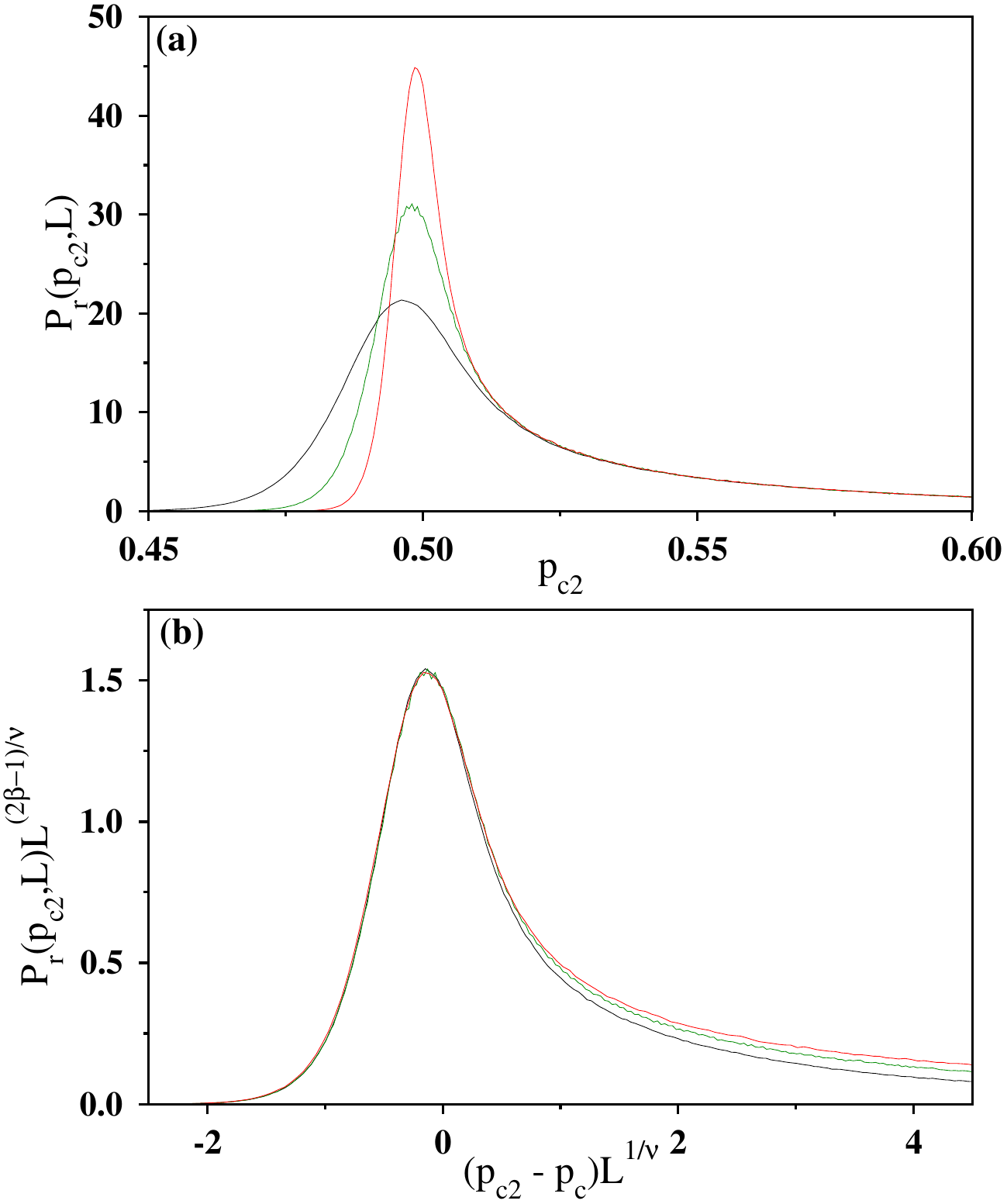}
   \caption{ Simulation result: (a) The probability distribution $P_r(p_{c2},L)$ of the 
   percolation threshold $p_{c2}$ of connecting two anchor points has been plotted 
   against $p_{c2}$ for $L$ = 128 \rev{(black or lower curve)}, 256 \rev{(green or middle curve)}, 512 \rev{(red or upper, most peaked curve)}. (b) The scaling plot 
   of the probability distribution $P_r(p_{c2},L)L^{(2\beta-1)/\nu}$ against 
   $(p_{c2}-p_c)L^{1/\nu}$ with $\beta=5/36, \nu=4/3$ with $p_c=1/2$. \rev{Bottom to top $L = 128$, 256, and 512.}
   }
   \label{fig:Prpc}
\end{figure}
%-----------------------------------------------------------------------------------

%-----------------------------------------------------------------------------------
\begin{figure}[htbp] %  figure placement: here, top, bottom, or page
   \centering
   \includegraphics[width=3.0 in]{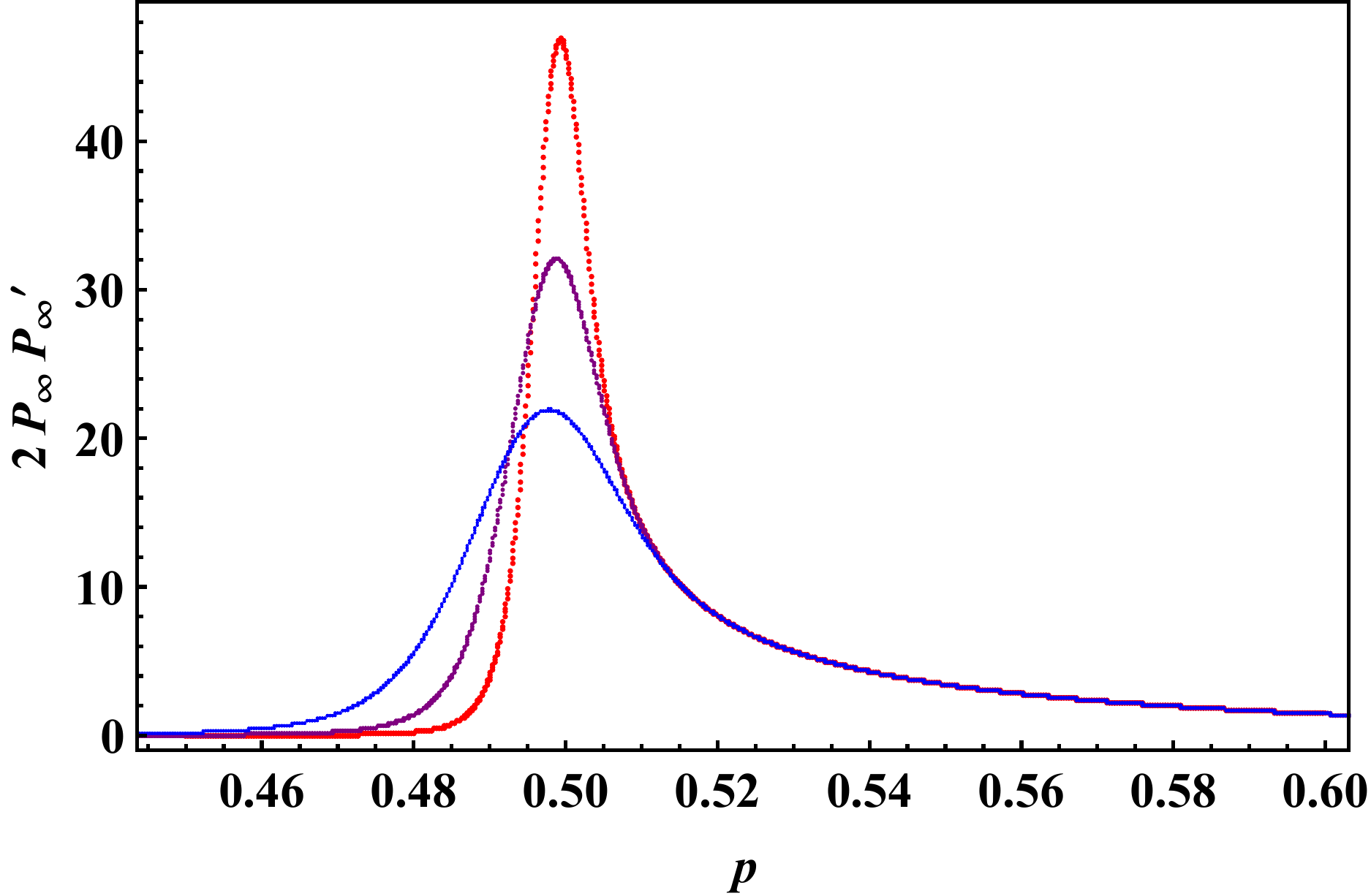} 
   \caption{ Plots of $\rev{P_r(p,L) = }2 P_\infty(p) P'_\infty(p)$ for $L = 128,$ 256 and 512, bottom to top at peaks. These 
   are the analogous curves as given in Fig.\ \ref{fig:Prpc}(a), calculated  
   from  $P_\infty(p)$ rather than by direct simulation \rev{of $P_r$.
   Note here $p$ is equivalent to $p_{c2}$ used in Fig.\ \ref{fig:Prpc}.}} 
   \label{fig:mannaP2p}
\end{figure}
%-----------------------------------------------------------------------------------

%-----------------------------------------------------------------------------------
\begin{figure}[htbp] %  figure placement: here, top, bottom, or page
   \centering
   \includegraphics[width=3.0in]{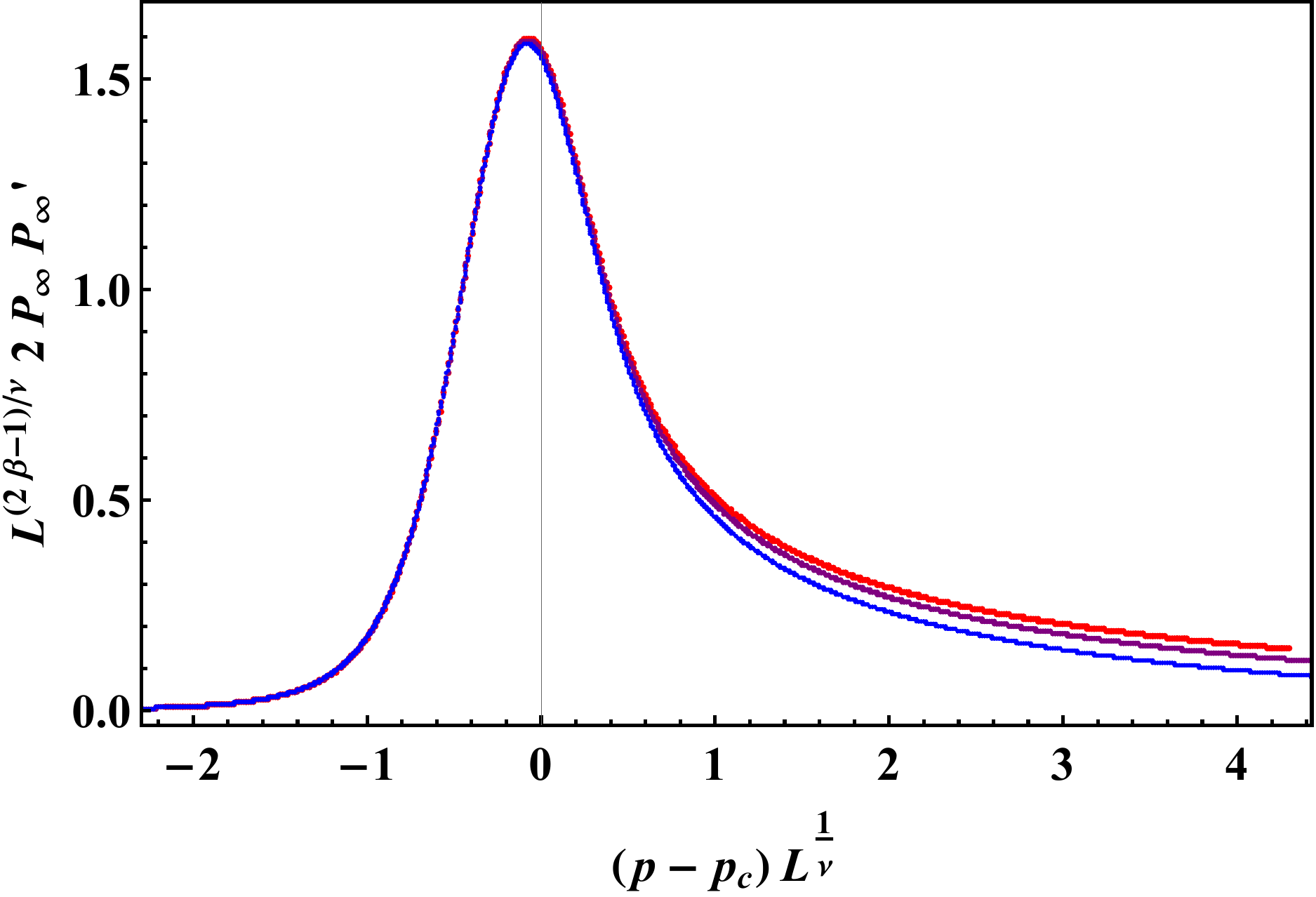} 
   \caption{ Scaling plot of $L^{(2 \beta-1)/\nu} 2 P_\infty(p) P'_\infty(p)$ 
   vs.\ $(p-p_c)L^{1/\nu}$ for $L = 128,$ 256 and 512 (bottom to top). The curves collapse well 
   to a universal curve, except for the tail for large $(p-p_c)L^{1/\nu}$ which 
   represents the non-scaling part of this quantity.  \rev{This plot is comparable with Fig.\ \ref{fig:Prpc}(b), here evaluated through $P_\infty$ rather than the direct measurement of $P_r$}.  Here $p$ is equivalent to $p_{c2}$ in Fig.\ \ref{fig:Prpc}.}    
   \label{fig:ScalingPlotManna}
\end{figure}
%-----------------------------------------------------------------------------------

\section{Simulation methods}
\label{sec:methods}

\subsection{Simulation method to find $\overline p_{ck}$}

      We carried out computer simulations of these processes on systems of size $L \times L$ 
   for bond 
   percolation, with periodic boundary conditions. For the case $k = 1$, we consider $L$ odd and 
   add bonds until the center point connects to the boundary for $p = p_{c1}$. Repeating this 
   process many times, we average the values of $p_{c1}$ to find $\overline p_{c1}$. For $k$ = 2, 3 and 4, we 
   consider periodic $L \times L$ systems with $L = 2^n$, $n = 5, 6, \ldots 12$. For $k = 2$ we 
   consider the connectivity between a point at the origin (0,0) and a point at $(0,L/2)$. For $k = 3$, 
   the connectivity between the three points (0,0), $(L/2,0)$, and $(0,L/2)$, and for $k = 4$, the 
   connectivity between the four points (0,0), $(L/2,0)$, $(0,L/2)$, and $(L/2,L/2)$ is considered.  \rev{Note that for $k = 3$, the three points are the vertices of a 
   right triangle rather than an equilateral triangle, so the distances between pairs of
   points are not identical, but this is not important --- all that matters is that the
   three points are relatively far apart from each other. }
   The average value of $p$ at the first connection gives $\overline p_{ck}$.

      It is  clear from Eq.\ (\ref{eq:Pransatz}) that the values of the thresholds $\overline p_{ck}$ should
   depend only on the value of $k$, and not on the actual distribution of the $k$ points. We
   have numerically verified this issue for $k=2$ by randomly distributing these two points on the 
   lattice for every configuration. Our simulation results show that the values of $\overline p_{c2}$ remain unchanged.

      We also studied the average $p$ at which the origin connects to point $x = 1$, $x = 2$, \ldots, 
   $x = L/2$ and $y = 0$ \rev{for systems of different $L$}. We discovered that $p_{c2}(x)$ does not noticeably depend upon $L$  as along as $x \ll L$, 
   indicating that the size of the system is \rev{unimportant} for shorter-range connections.

\subsection{Simulation method to find $P_\infty$}

      To test the conjecture relating $\overline p_{ck}$ to $P_\infty$, we carried out 
   measurements of $P_\infty(p)$ using the method of Newman and Ziff (NZ) 
   \cite{NewmanZiff00,NewmanZiff01}, which involves adding bonds one at a time to the 
   system and using the union-find procedure to merge clusters and keep track of the 
   cluster distribution. This method allows one to effectively measure a quantity 
   $Q(p)$ (such as $P_\infty(p))$ for all values of $p$ in a single simulation. In this 
   method, one first determines the ``microcanonical'' $Q_n$ (here $P_{\infty,n}$) when 
   exactly $n$ bonds have been placed down, and then determines the ``canonical''
   $Q(p)$ (here $P_\infty(p)$) by carrying out a convolution with the binomial distribution 
   $B(N,n,p)={N \choose n} p^n (1-p)^{N-n}$:
\begin{equation}
Q(p) = \sum_{n=0}^N {N \choose n} p^n (1-p)^{N-n} Q_n
\label{eq:convolution}
\end{equation}
   where $N$ is the total number of bonds in the system, in this case $2L^2$. For large 
   systems, the differences between the microcanonical $Q_n$ with $n=p N$ and $Q(p)$ are  
    small, except for regions of high curvature or second derivative, but the convolution serves a 
   further purpose of smoothing out the data, and connecting it with a continuous curve, 
   rather than the discrete values $p = 1/N, 2/N, \ldots$. To integrate $P_\infty(p)$ 
   (as required for $\overline p_{c1}$ according Eqs.\ (\ref{eq:pck}) or (\ref{eq:pckalt})), 
   one can just as well sum the microcanonical values, because of the identity \cite{ZiffNewman02}
\begin{eqnarray}
\int_0^1 Q(p) dp &=& \sum_{n=0}^N
  {N \choose n}  Q_n \int_0^1 p^n (1-p)^{N-n} dp  \cr
&=& \frac{1}{N+1}  \sum_{n=0}^N Q_n
\end{eqnarray} 
   Likewise it follow that
\begin{eqnarray}
\int_0^1 p Q(p) dp
&=& \frac{1}{(N+1)(N+2)}  \sum_{n=0}^N (n+1) Q_n
\end{eqnarray} 
   To integrate $[Q(p)]^k = [P_\infty(p)]^k$ for $k>1$ with respect to $p$, it is most 
   straightforward to first carry out the convolution to find $P_\infty(p)$, and then 
   numerically integrate the $[P_\infty(p)]^k$ at equally spaced values of $p$.

Derivatives of $Q(p)$ can also be found directly from the $Q_n$ \cite{ZiffNewman02}:
\begin{eqnarray}
Q'(p) &=& \sum_{n=0}^N  {N \choose n}   \rev{Q_n} \frac{d}{dp} \left( p^n (1-p)^{N-n}\right)  \cr
&=& \frac{1}{p(1-p)}  \sum_{n=0}^N  (n-Np) {N \choose n} ( p^n (1-p)^{N-n})Q_n \cr
&= &\frac{\langle( n - N p)Q_n \rangle }{p(1-p)} 
\label{eq:derivative}
\end{eqnarray}
and likewise

\begin{eqnarray}
 &\textstyle{Q''(p)}&=\textstyle{ \frac{\langle n^2 Q_n \rangle    -(2(N-1)p+1) \langle n Q_n \rangle + N(N-1)p^2\langle Q_n \rangle}{p^2(1-p)^2} \hspace{1 em}} \cr 
&\textstyle{=}&\textstyle{\hspace{-1.1em} \frac{\langle (n-Np)^2 Q_n\rangle+(2 p - 1)   \langle (n - N p) Q_n\rangle - N p (1-p)\langle Q_n \rangle }{p^2(1-p)^2} }
\label{eq:secondder}
\end{eqnarray}
%\begin{widetext}
%\begin{eqnarray}
% &Q''(p)&= \frac{\langle n^2 Q_n \rangle    -(2(N-1)p+1) \langle n Q_n \rangle + N(N-1)p^2\langle Q_n \rangle}{p^2(1-p)^2} \cr 
%&=&\frac{\langle (n-Np)^2 Q_n\rangle+(2 p - 1)   \langle (n - N p) Q_n\rangle - N p (1-p)\langle Q_n \rangle }{p^2(1-p)^2} 
%\label{eq:secondder}
%\end{eqnarray}
%\end{widetext}
where the averages are over the binomial distribution $B(N,n,p)$.  Note that in Ref.\ \cite{ZiffNewman02},
there is a typo in Eq.\ (32) for $Q''(p)$, in which the last term should have the factor $(N-n-1)$ rather than $(N-n+1)$.

%-----------------------------------------------------------------------------------
\begin{figure}[htbp] %  figure placement: here, top, bottom, or page
   \centering    
   \includegraphics[width=3 in]{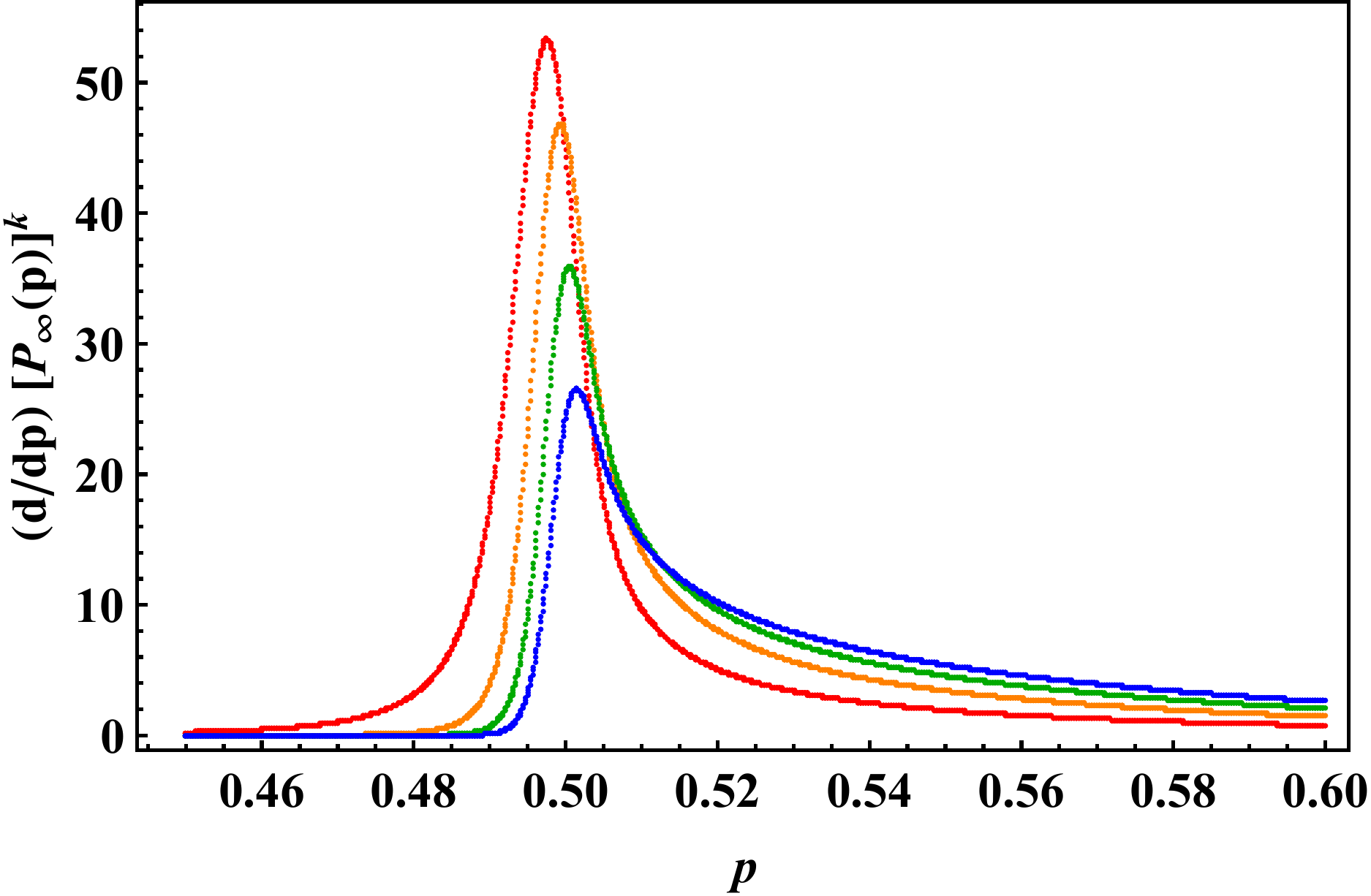} 
   \caption{ A plot of $(d/dp) [P_\infty(p)]^k$ vs.\ $p$ for $k = 1$ (red, highest peak)), $k = 2$ 
   (orange, second-highest peak), $k=3$ (green, second-lowest peak),
   and $k=4$ (blue, lowest peak) for a system with $L = 512$, calculated 
   from the results of the numerical simulations of $P_\infty(p)$, including using Eq.\ (\ref{eq:derivative}) to find $P'_\infty(p)$. The 
   estimates of $\overline p_{ck}$ are the means of these distributions according to 
   Eq.\ (\ref{eq:pck1}), and it can be seen that the distribution spreads to the 
   right as $k$ increases, yielding larger values of $\overline p_{ck}$. To find the 
   accurate values of $\overline p_{ck}$, one has to  consider systems of different $L$ 
   and take the limit that $L \to \infty$, although \rev{the change is small} for systems 
   larger than $L = 512$.  Note that the distribution is broad and the large fluctuations in the 
   individual values of $p_{ck}$ persist as $L \to \infty$.}
   \label{fig:Pk}
\end{figure}
%-----------------------------------------------------------------------------------

      To find $P_\infty(p)$ we simulated $10^7$ samples each for $L = 64, 128, 256$ 
   and 512 on $L \times L$ periodic systems, saving the $2 L^2$ microcanonical values 
   of $s_\mathrm{max}$ in a file. For the largest system $L = 512$, the simulations took 
   several days on a laptop computer. Then we used a separate program to read the 
   files and calculate $P_\infty(p) = s_\mathrm{max}/L^2$ for $10^4$ points $p = 0, 0.0001, 
   \ldots, 1.0000 $ using the convolution (\ref{eq:convolution}).  We also calculated 
   $P'_\infty(p)$ and $P''_\infty(p)$ using the formulas of Eqs.\ (\ref{eq:derivative}) 
   and (\ref{eq:secondder}).   We used the recursive method described in Ref.\ 
   \cite{NewmanZiff01} to calculate the binomial distribution for each $p$. To find the 
   integrals of $[P_\infty(p)]^k$ \rev{for Eqs.\ (\ref{eq:pck}) and (\ref{eq:pckalt})}, we carried out numerical integration of the $10^4$ 
   points using the trapezoidal rule (namely counting the two endpoints with \rev{relative} weight 1/2 
   and all other points with weight 1).  We compared some of the integrals using $10^3$ 
   and $10^5$ points and did not find significant difference in the results, and used 
   $10^4$ values of $p$ in our calculations.

\section{Results}
\label{sec:results}

\rev{Figure \ref{fig:Prpc}(a) shows the probability distribution $P_r(p_{c2},L)$ of the 
  percolation threshold $p_{c2}$ of connecting two anchor points, from direct
  measurements.  Note $p_{c2}$ is the value of $p$ at which the connection first
  takes place in a given sample, as opposed to $\overline p_{c2}$ which is the 
  average value over many samples.   Figure \ref{fig:Prpc}(b)  shows a scaling plot of the data, using the 
  scaling implied in Eq.\ \ref{eq:Prscaling}.}

\rev{Figure \ref{fig:mannaP2p} shows the predicted behavior of $P_r$ from 
the ansatz of Eq.\ \ref{eq:Pransatz}, using the simulation results of $P_\infty$ rather than measuring
 $P_r$ directly.  These curves can be compared with those of Fig.\  \ref{fig:Prpc}(a), and 
 the two can be seen to agree.}

\rev{Figure \ref{fig:ScalingPlotManna} shows the predicted scaling behavior of $P_r$ from
the ansatz of Eq.\ \ref{eq:Pransatz}, and the results can be seen to be similar to the scaling plot of the directly measured $P_r$ given in Fig.\ \ref{fig:Prpc}(b).}

Figure \ref{fig:Pk} shows plots of the predicted distributions of the probabilities of first connection,
 $(d/dp) [P_\infty(p)]^k$, for $k = 1$, 2, 3, and 4, based upon measurements $P_\infty(p)$, for a system
of $L = 512$.   As can be seen, the distributions are broad, meaning that the
 thresholds we find $p_{ck}$ have large
 fluctuations from system to system and persist as $L \to \infty$. 

%-----------------------------------------------------------------------------------
\begin{figure}[htbp]
   \centering
   \includegraphics[width=3 in]{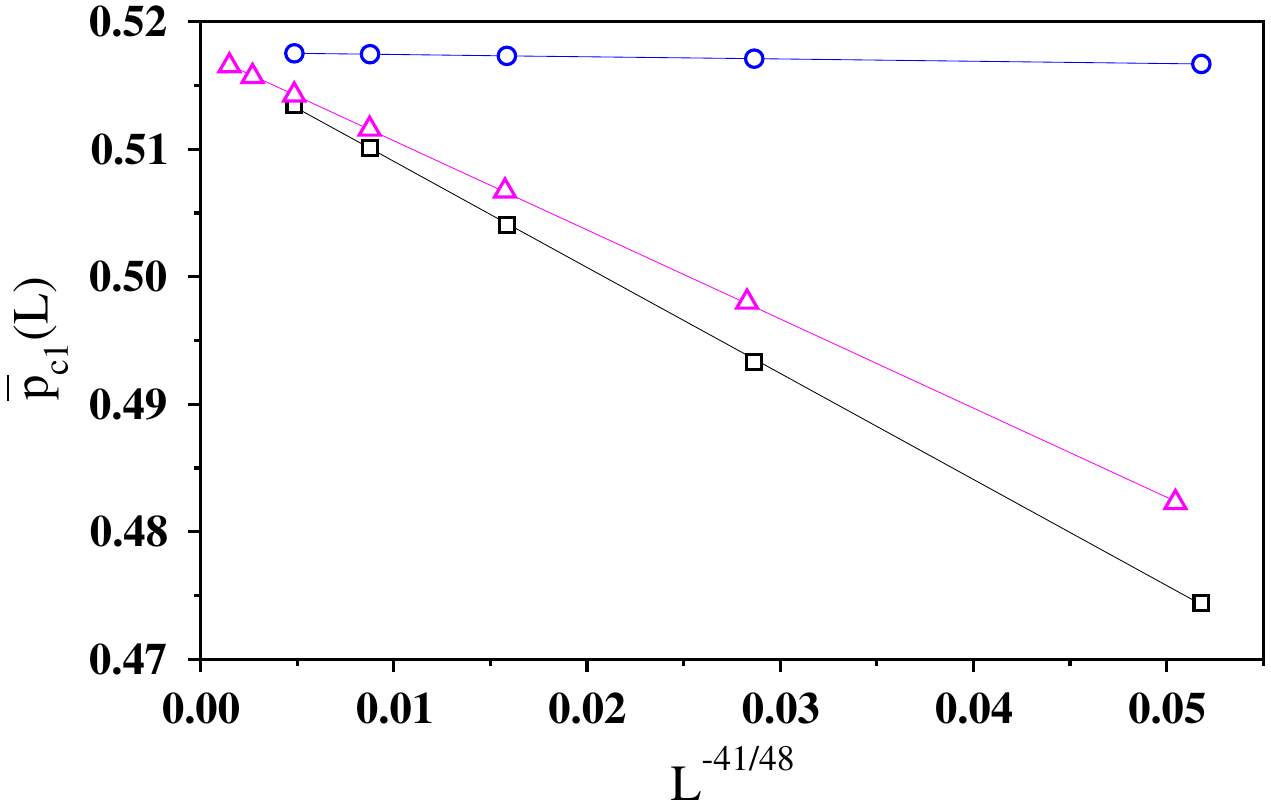} 
   \caption{
   Values of $\overline p_{c1}(L)$ found from  simulations of connections of 
   a point at the center to the boundary of an $(L+1) \times (L+1)$ square system 
   (triangles), by integrating $P_\infty(p)$ on $L \times L$ periodic systems 
   using Eq.\  (\ref{eq:pck}) for $k = 1$  (squares), and by integrating 
   $P_\infty(p)$ using Eq.\  (\ref{eq:pckalt}) (circles). The estimates are all 
   plotted vs.\  $L^{-41/48}$ according to the prediction of Eq.\  (\ref{eq:nuk}).  
   The equations of the linear fits through the points are $\overline p_{c1} = 
   a + b L^{-41/48} $ with $a=0.55520$, $b=-0.33805$ (squares), 
   $a=0.55532$, $b=-0.27962$ (triangles), $ a=0.55530$, $b=-0.06323$ (circles).
   }
   \label{fig:pc1}
\end{figure}
%-----------------------------------------------------------------------------------

      In Fig.\ \ref{fig:pc1} we plot estimates for $\overline p_{c1}$ found from direct simulations with the point 
   in the center of an $(L+1) \times (L+1)$ system, for $L = 64, 128, \ldots, 4096$, and secondly using the 
   formulas of Eqs.\  (\ref{eq:pck}) and  (\ref{eq:pckalt}) \rev{for $k=1$} based upon $P_\infty(p)$.  The data are 
   plotted based on the predicted scaling $L^{-41/48}$ \rev{from Eq.\ (\ref{eq:nuk})}.  We do not expect that the values of $\overline p_{c1}(L)$ 
   would be the same for finite $L$ from the two methods (direct simulation and via $P_\infty$); however, we expect that the extrapolation as 
   $L \to \infty$ should be the same, because in that limit the probability the point connects to the boundary 
   should exactly be the probability the point belongs to the largest cluster, namely $P_\infty$.  Furthermore, 
   we expect the two estimates of $\overline p_{c1}$ should scale with $L$ with the same exponent 
   $1/\nu_1 = 41/48$, and indeed that plot confirms that expectation.  The two different approaches suggest a 
   threshold of $\overline p_{c1} = 0.51749(5)$.

\begin{figure}[htbp] %  figure placement: here, top, bottom, or page
   \centering
      \includegraphics[width=3 in]{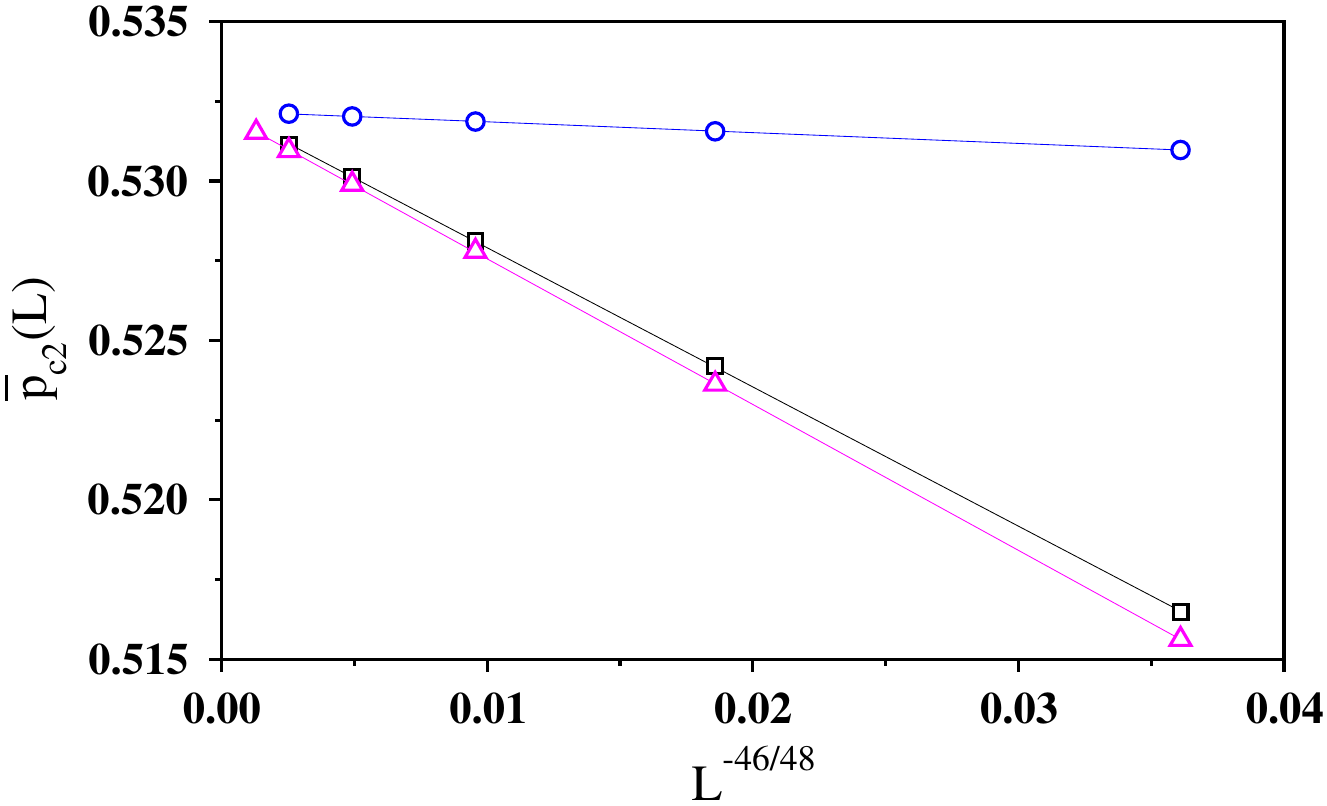} 
   \caption{ Values of  $\overline p_{c2}(L)$ found from direct simulations on an $L 
   \times L$ periodic system with the two points at (0,0) and (0,$L/2$) (triangles),
   and the predictions from Eqs.\ (\ref{eq:pck}) (squares) 
   and (\ref{eq:pckalt}) (circles) based upon measurements of $P_\infty(p)$ on an $L \times L$ 
   periodic system, all plotted as a function of $L^{-46/48}=L^{-23/24}$ as 
   predicted by Eq.\ (\ref{eq:nuk}). Here  $L = 32$, 64, 128, 256, and 512 for 
   the upper two sets of data, and also $L = 1024$ for the lower set.
   }
   \label{fig:pc2}
\end{figure}

\begin{figure}[htbp] %  figure placement: here, top, bottom, or page
   \centering
   \includegraphics[width=3 in]{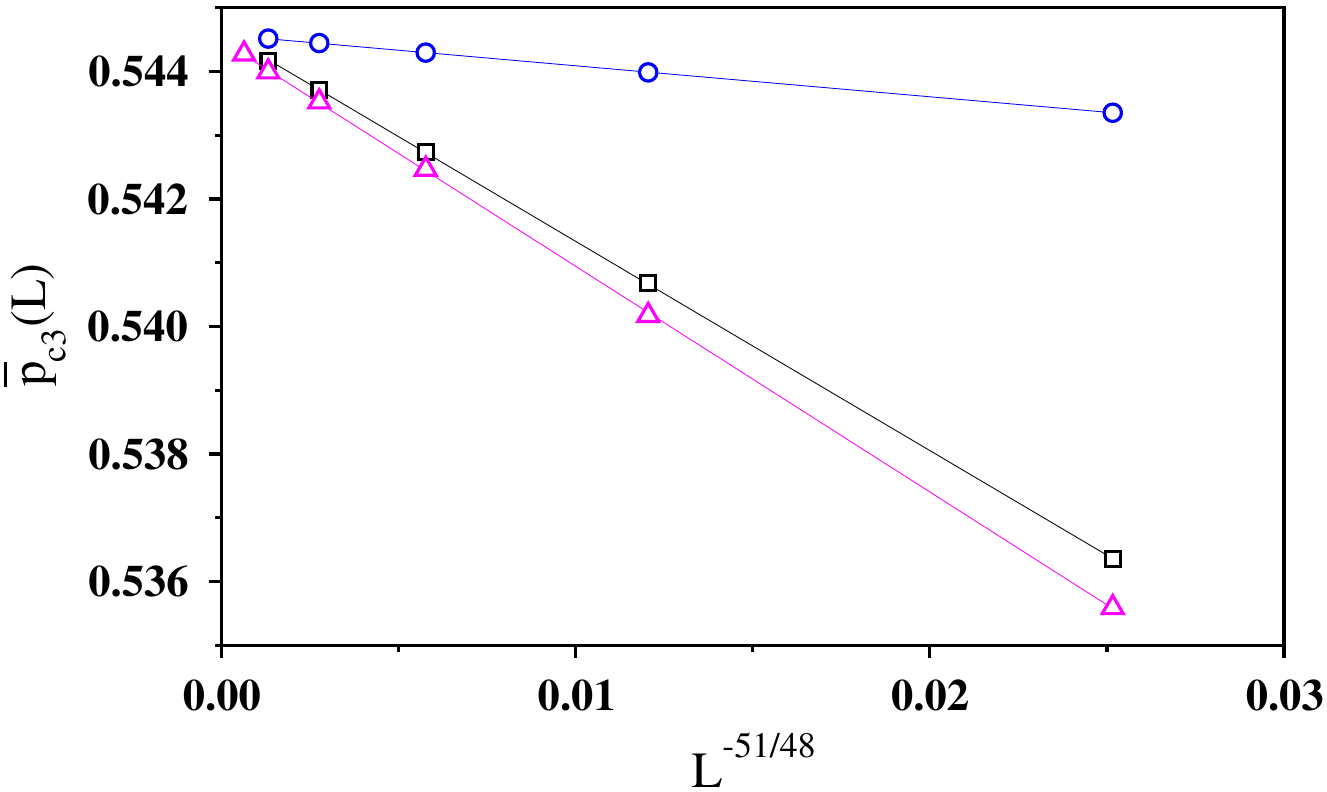} 
   \caption{ Values of  $\overline p_{c3}(L)$ found from direct simulations on an 
   $L \times L$ periodic system with the three points at (0,0), $(0,L/2)$ and ($L/2,0$)
   (triangles), and the predictions from Eqs.\ (\ref{eq:pck}) (squares) and (\ref{eq:pckalt})
    (circles) based upon measurements of $P_\infty(p)$ on an 
   $L \times L$ periodic system, all plotted as a function of $L^{-51/48}=L^{-17/16}$ 
   as predicted by Eq.\ (\ref{eq:nuk}).
   }
   \label{fig:pc3}
\end{figure}

\begin{figure}[htbp]
   \centering
   \includegraphics[width=3 in]{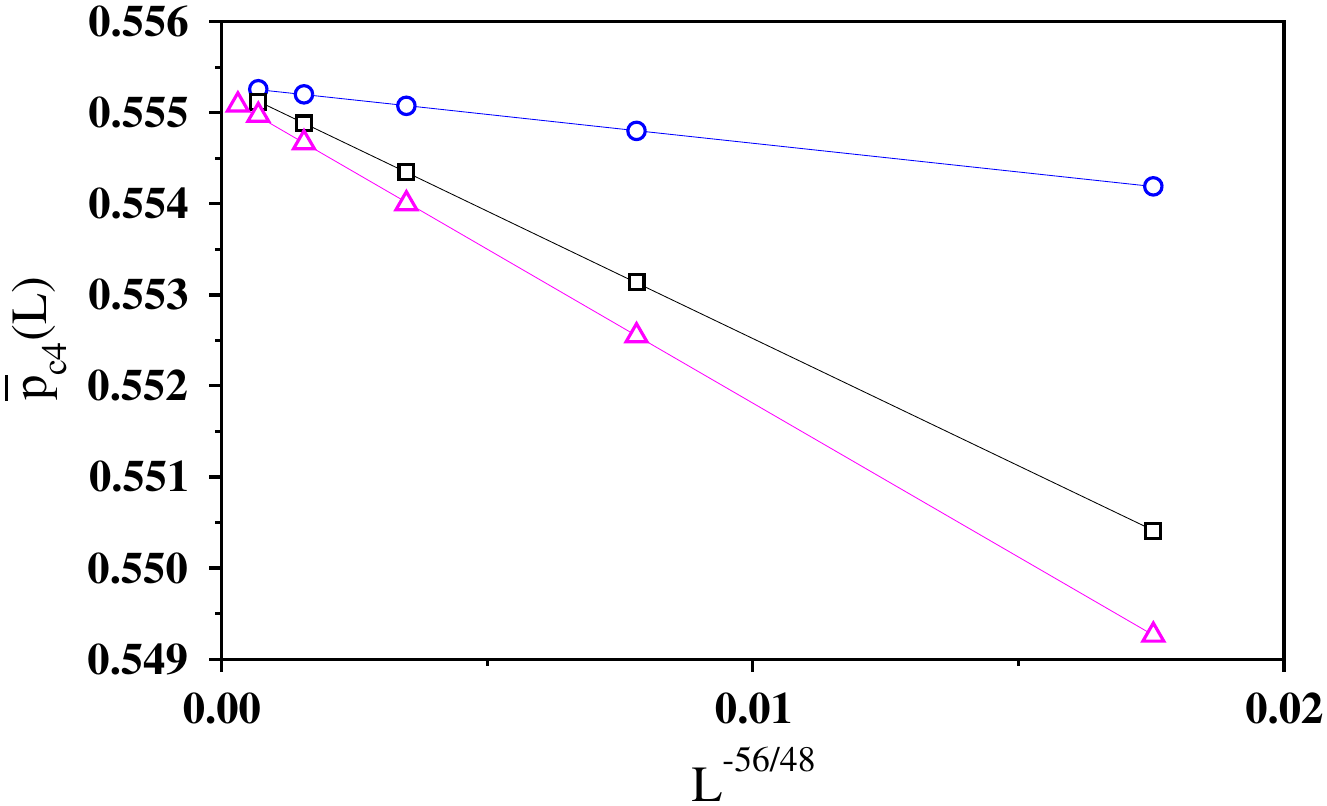} 
   \caption{ Values of $\overline p_{c4}(L)$ found from direct simulations on an 
   $L \times L$ periodic system with the four points at the corners of a square of 
   length $L/2$ (triangles), and the predictions based upon $P_\infty(p)$ 
   using Eq.\ (\ref{eq:pck}) (squares) and Eq.\ (\ref{eq:pckalt}) (circles), 
   both based upon  measurements of $P_\infty(p)$ on an $L \times L$ periodic system. All data are 
   plotted as a function of $L^{-56/48}=L^{-7/6}$ as predicted by Eq.\ (\ref{eq:nuk}).  
   Lines show linear fits through the data. It can be seen that estimates based upon 
   Eq.\ (\ref{eq:pckalt}) exhibit the fastest convergence with system size.
  }
   \label{fig:pc4}
\end{figure}

      It can clearly be seen that the estimate based upon (\ref{eq:pckalt}), which assumes $P_\infty(p,L)=0$ 
   for $p < p_c$, converges much more quickly than the estimate based upon (\ref{eq:pck}). On a more expanded 
   scale, the convergence to this estimate is also shown to obey the $L^{-41/48}$ scaling, but is not shown 
   here.  The results for $k = 2$, 3 and 4 are shown in Figs.\ \ref{fig:pc2}, \ref{fig:pc3}, and \ref{fig:pc4}.
  Our values of $p_{ck}$ are given in Table \ref{tab:01}.

%-----------------------------------------------------------------------------------
%-----------------------------------------------------------------------------------

%-----------------------------------------------------------------------------------
%-----------------------------------------------------------------------------------

%-----------------------------------------------------------------------------------
%-----------------------------------------------------------------------------------

\section{Correlations}
\label{sec:correlations}

We also considered a related question for two- and three-point correlations.
Studying this problem sheds light on the correlations that occur in the system in the critical vs.\ the post-critical regime where 
the connectivity between the anchor points mainly occurs.
 
In \cite{SimmonsZiffKleban09,DelfinoViti10} the following ratio was considered:
\begin{equation}
\rev{R(p)} = \frac{P(r_1,r_2,r_3)}{\sqrt{P(r_1,r_2) P(r_1,r_3) P(r_2,r_3)}}
\label{eq:Rratio}
\end{equation}
where $r_1$, $r_2$ and $r_3$ are three points in the system, $P(r_i,r_j)$ is the probability that points $r_i$ and $r_j$ connect, and $P(r_1,r_2,r_3)$ is the probability that all three points connect. 

\begin{figure}[htbp] %  figure placement: here, top, bottom, or page
   \centering
   \includegraphics[width=3 in]{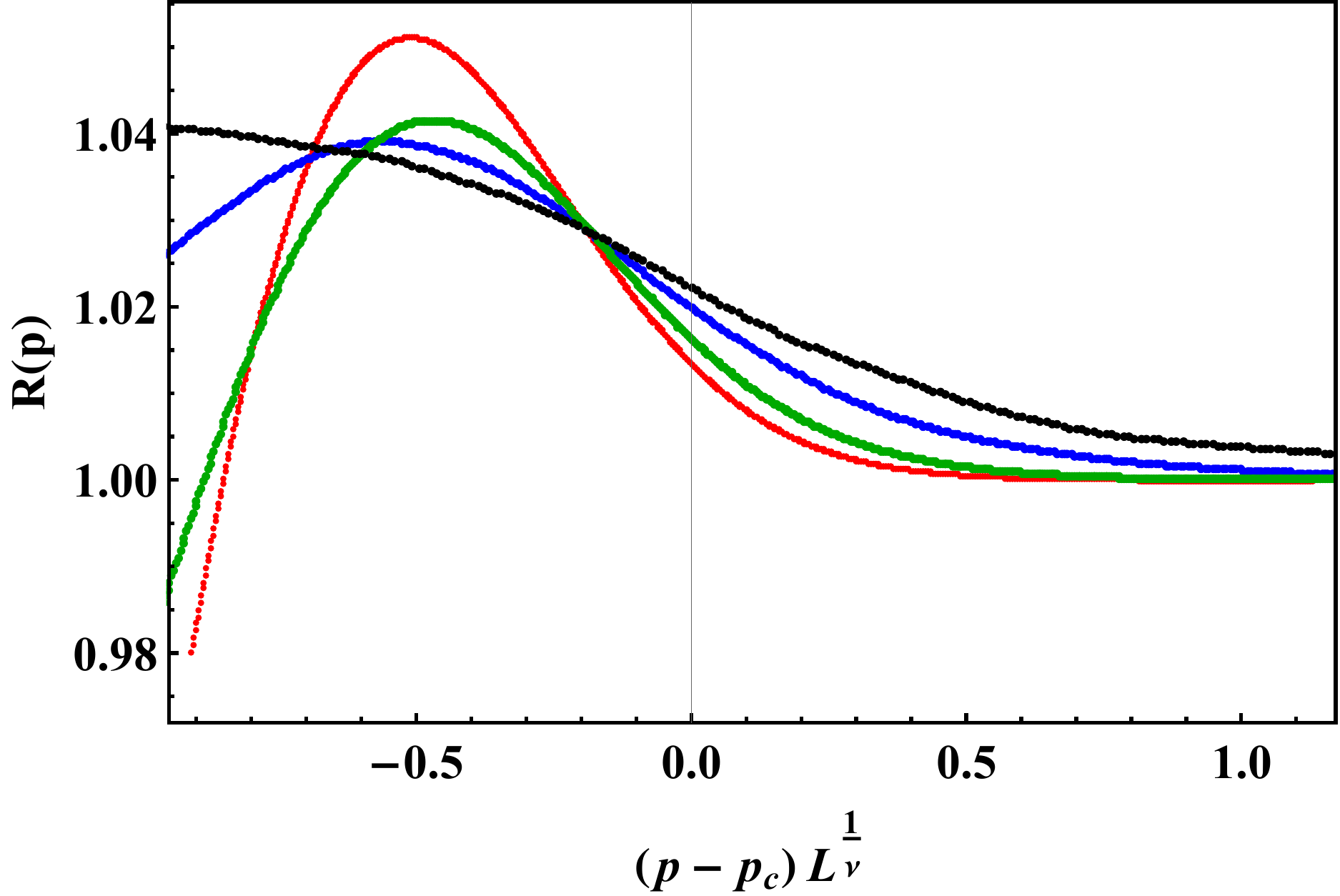}
  \caption{\rev{(color online) Scaling plot of $R(p)$ vs.\ $(p-p_c) L^{1/\nu}$ where $R(p)$ is given in Eq.\ (\ref{eq:Rratio}), for three points at positions ($0,0$), ($0,L/n$), and ($L/n,0$), for $n = 2$ (red, the  curve with the highest peak), $n = 4$, (green, the curve with the second-highest peak), $n = 8$ (blue, the curve with the third-highest peak), and $n = 16$ (black, the curve that does not reach a peak in this interval), for $L = 64$, 128, 256, and 512 respectively. Other runs show that there is a small  $L$-dependence on the results, but the main variation is due to $n$.  At $p = p_c = 1/2$, the value of $R(p_c)$ approaches the theoretical value $C_1 = 1.022$  \cite{SimmonsZiffKleban09,DelfinoViti10} as $n$ gets large, in which case the three points are close together compared to the size of the system.  The meaning of the crossing point for $(p-p_c)L^{1/\nu} \approx -0.2$ is unclear, and may not be maintained for larger systems.}}
 \label{fig:Rratio}  
\end{figure}

      This ratio has previously been studied, to our knowledge, only at $p = p_c$, where the value of $R(p_c)$ approaches the value $C_1 = 1.0220$ 
   when the three points are far separated and the system size is infinite.  This value of $C_1$ was first observed numerically in 
   \cite{SimmonsZiffKleban09} and then derived analytically from conformal field theory in \cite{DelfinoViti10}. The fact that this ratio is 
   unequal to 1 implies a correlation between the three points in the system.   \rev{If we make the assumption that $P(r_i,r_j) = 
   P_\infty(p)^2$ and $P(r_1,r_2,r_3) = P_\infty(p)^3$, which we expect to be the case for $p > p_c$, then we would have $R = 1$.} At $p_c$, where 
   the infinite cluster does not span throughout the system, one would not expect this to be valid and indeed $R(p) \ne 1$, although \rev{it turns out quite close to 1.}

      Here we consider the three points in a right triangle, ($0,0$), ($0,L/n$), and ($L/n,0$), in an $L \times L$ periodic system, for $n = 2,$ 4 and 8 . As $n$ increases 
  for large $L$ (that is, as the separation of the three points is small compared to the size of the system), $R(p_c)$ approaches the value 
   $C_1$. Using the NZ method, we were able to calculate $R(p)$ as a function of $p$ after executing a microcanonical simulation where we found the 
   $P(r_i,r_j)$ and $P(r_1,r_2,r_3)$ as a function of the number of bonds added.  We then carried out the convolution to the canonical ($p$-dependent) functions for all $P$'s 
   separately, and  calculated $R(p)$ according to Eq.\ (\ref{eq:Rratio}).  The results are shown in Fig.\ \ref{fig:Rratio}.

      As can be seen, at $p = p_c$, $R(p_c)$ approaches $C_1$ as $n$ increases (in which case the points get closer together compared to the size of the system). 
In the limit that $L \to \infty$, $R(p)$  \rev{evidently} becomes a discontinuous function \rev{of $p$}, with $R(p_c)=C_1$ for 
   $p = p_c$, and $R(p)=1$ for $p>p_c$.  \rev{The behavior for $p < p_c$ is not clear.}  Notice in Fig.\ \ref{fig:Rratio} that there is a maximum for $R(p)$ in finite systems at $z = (p-p_c) L^{1/\nu} 
   \approx -0.5$, meaning for some values of $p < p_c$, $R(p)$ is greater than the value at $p_c$. However, it is not clear what the behavior is as 
   $n\to \infty$ (for large $L$); it is possible that the peak for negative $z$ disappears and the peak occurs only at $z = 0$ or $p = p_c$.  \rev{The behavior for $p<p_c$ needs further investigation.}

      At the point $p_{c3} \approx 0.5445$ where three points first connect, it can be seen that $R(p)$ approaches 1, since that would correspond to $(p-p_c) L^{1/\nu}$ going 
   to infinity as $L$ goes to infinity. This result reiterates that at the places where multiple points connect, there are no correlations among 
   connections between different pairs of \rev{widely separated} points. 

%-----------------------------------------------------------------------------------
\begin {table}[t]
\caption{
         Our best estimates for the extrapolated values of the average percolation thresholds 
         $\overline p_{ck}$ from direct measurements (second column) and from $P_\infty$ 
         via Eqs.\ (\ref{eq:pck}) and  (\ref{eq:pckalt}) for different values of $k$.  The averages
         of these values are quoted in the abstract.
         }
\begin{tabular}{llll} \\ \hline
$k$   \hspace*{1.0 cm}& $\overline p_{ck}$ measured     \hspace*{1.0 cm}& Eq.\ (\ref{eq:pck})   \hspace*{1.0 cm}& Eq.\ (\ref{eq:pckalt})    \\ \hline
1    & 0.51749(5)   & 0.51761(3)  & 0.51755(3)    \\
2    & 0.53212(5)   & 0.53220(3)  & 0.53226(3)    \\
3    & 0.54450(5)   & 0.54458(3)  & 0.54461(3)    \\
4    & 0.55520(5)   & 0.55530(3)  & 0.55531(3)    \\ \hline
\end{tabular}
\label {tab:01}
\end {table}
%-----------------------------------------------------------------------------------

\section{Discussion}
\label{sec:discussion}

      We have shown that exploring the average value of the probability $p$ of bond occupation at which a certain number of 
   separated points first connect leads to a new set of average thresholds. The distribution of 
   the values of $p$ is broad, so that this threshold is not sharp as in the 
   usual case of thresholds in percolation. For example, the median rather than the mean 
   of the distribution would give a different value. We have shown that the values can 
   be related to $P_\infty(p)$, and confirm this relation by simulation. From this theory 
   it is apparent that while the percolation thresholds $\overline p_{ck}$ indeed depend on
   the number $k$ of points, their values are robust with respect to the actual spatial
   distribution of the $k$ points. For example, the $k$ points may either be symmetrically
   placed on the lattice or, they can be randomly distributed (for $L \to \infty$).

      This work suggests further research in a variety of areas. It might be interesting 
   to study these thresholds in higher dimensions, where the relations to $P_\infty(p)$ 
   in Eqs.\ (\ref{eq:pck}) and (\ref{eq:pckalt}), and the scaling in (\ref{eq:nuk}) 
   (but with $\nu$ and $\beta$ being the three-dimensional result) should still hold, 
   for connections to points as we considered here. Furthermore, connections between 
   higher-dimensional objects (lines, surfaces, ...) can also be considered. One question 
   to consider is whether the thresholds continue to have broad distributions as found 
   here, and how those thresholds scale with $L$.

      With respect to the correlations $R(p)$, one can consider a point in the center of a 
  \rev{surface of }a cylinder (that is, the center of a square with periodic b.c.\ in one direction), and find the 
   probability of connecting the center to one boundary or to both boundaries of the 
   cylinder. At $p_c$, the corresponding $R(p)$  should go to the value  $C_0 =  2^{7/2} 
   3^{-3/4} \pi^{5/2} \Gamma(1/3)^{-9/2} = 1.029 9268\ldots$ \cite{SimmonsZiffKleban09} 
   while the behavior away from $p_c$ has not been studied before. Likewise, similar 
   correlations in higher dimensions have not been studied. Many aspects of correlations 
   in percolation are yet to be explored.
   
   \section{Acknowledgment} The authors would like to thank Deepak Dhar for a careful reading
   and  constructive comments on the paper.

\bibliographystyle{unsrt}
\bibliography{bibliography19.bib}
\end{document}